\documentclass[12pt]{article}

\usepackage{scicite}
\usepackage{times}
\usepackage{amsmath}
\usepackage{graphicx}
\usepackage{gensymb}
\usepackage{bbold}
\usepackage{braket}
\usepackage{textgreek}
\usepackage{hyperref}
\usepackage{amssymb}

\newenvironment{sciabstract}{%
\begin{quote} \bf}
{\end{quote}}

\title{Metasurface-enabled compact, single-shot and complete Mueller matrix imaging}

\author
{Aun Zaidi,$^{1\ast}$ Noah A. Rubin,$^{1}$ Maryna L. Meretska,$^{1}$ Lisa Li,$^{1}$ \\ Ahmed H. Dorrah,$^{1}$ Joon-Suh Park,$^{1,2}$ Federico Capasso$^{1\ast}$\\
\\
\normalsize{$^{1}$Harvard John A. Paulson School of Engineering and Applied Sciences,}\\ \normalsize{Harvard University, Cambridge, Massachusetts 02138, USA}\\
\normalsize{$^{2}$Nanophotonics Research Center, Korea Institute of Science and Technology,}\\ \normalsize{Seoul 02792, Republic of Korea}\\
\\
\normalsize{$^\ast$To whom correspondence should be addressed; E-mail:  zaidi01@g.harvard.edu (A.Z.);}\\ \normalsize{capassso@seas.harvard.edu (F.C.)}
}

\date{}

\begin{document} 


\baselineskip24pt

\maketitle


\begin{sciabstract}
  When light scatters off an object its polarization, in general, changes -- a transformation described by the object's Mueller matrix. Mueller matrix imaging polarimetry is an important technique in science and technology to image the spatially varying polarization response of an object of interest, to reveal rich information otherwise invisible to traditional imaging. In this work, we conceptualize, implement and demonstrate a compact and minimalist Mueller matrix imaging system -- composed of a metasurface to produce structured polarization illumination, and a metasurface for polarization analysis -- that can, in a single shot, acquire images for all sixteen components of an objects’s spatially varying Mueller matrix. Our implementation, which is free of any moving parts or bulk polarization optics, should enable and empower applications in real-time medical imaging, material characterization, machine vision, target detection, and other important areas.  
\end{sciabstract}

% In setting up this template for *Science* papers, we've used both
% the \section* command and the \paragraph* command for topical
% divisions.  Which you use will of course depend on the type of paper
% you're writing.  Review Articles tend to have displayed headings, for
% which \section* is more appropriate; Research Articles, when they have
% formal topical divisions at all, tend to signal them with bold text
% that runs into the paragraph, for which \paragraph* is the right
% choice.  Either way, use the asterisk (*) modifier, as shown, to

Polarization, an important and fundamental property of light, is classically defined as the direction of oscillation of the light's electric field. Polarization, as a design degree of freedom and a source of additional information, has always been of great interest and utility to both fundamental science and technological innovation \cite{Snik2014,Damask2005, Goldberg2021}. The generation, manipulation, sensing, and imaging of polarization is of paramount importance because of the potential of polarization to reveal rich information about the physical properties of objects, materials and their structures, indiscernible to other optical techniques \cite{Tyo2006,Demos1997,Tyo1996}. Polarization can be represented by a four-component vector, known as a Stokes vector, and consequently, linear interaction of light with a sample can be described fully by a $4 \times 4$ matrix, known as a Mueller matrix (MM). Polarization transforming properties of an object, such as its retardance, diattenuation, polarizance and depolarization can all be directly computed from its MM \cite{Chipman2019}. MM imaging polarimetry captures images of the spatially varying MM elements of a sample or object of interest, providing the most complete image of an object's polarization properties, and revealing information which would otherwise be invisible to traditional intensity-only imaging \cite{Azzam2016}. 

MM imaging has found important applications in different fields of science and technology, most noticeably in biology and medicine \cite{Tuchin2006, Alali2015}. Some examples include the use of MM imaging for disease diagnostics \cite{Dubreuil2012}, retinal imaging \cite{Weinreb1990,Dreher1992}, glucose sensing \cite{Westphal2016}, bacteria detection \cite{Svensen2011,Li2017}, identification of malignant tissues \cite{Ghosh2011}, differentiation between healthy and cancerous tissues \cite{Baldwin,Antonelli2010,Pierangelo2013} (cancerous tissues, for instance, exhibit different depolarization and retardance signatures than healthy tissues), and for discerning between different types and stages of cancers \cite{Novikova2012}. The use of MM imaging has also been explored in areas such as ellipsometry \cite{Azzam1997,Liu2015}, oceanic sciences \cite{Kokhanovsky2003}, turbidity \cite{Kattawar2003}, and target detection \cite{Tyo1996,Azzam2016}. Despite the obvious advantages and utility of MM imaging, its widespread adoption has been hampered by the relative complexity, bulk, cost and hardware limitations of its existing implementations. In our work, using metasurface optics, we propose and implement a first-of-its-kind compact, single-shot and complete MM imaging system that overcomes the limitations of existing MM imaging solutions, and provides a viable pathway for the widespread adoption of MM imaging to empower applications requiring advanced imaging or sensing modalities. 

\subsection*{Conventional vs. Metasurface Polarization Optics}

\begin{figure*}[t!]
	\centering
    \includegraphics[width=\textwidth]{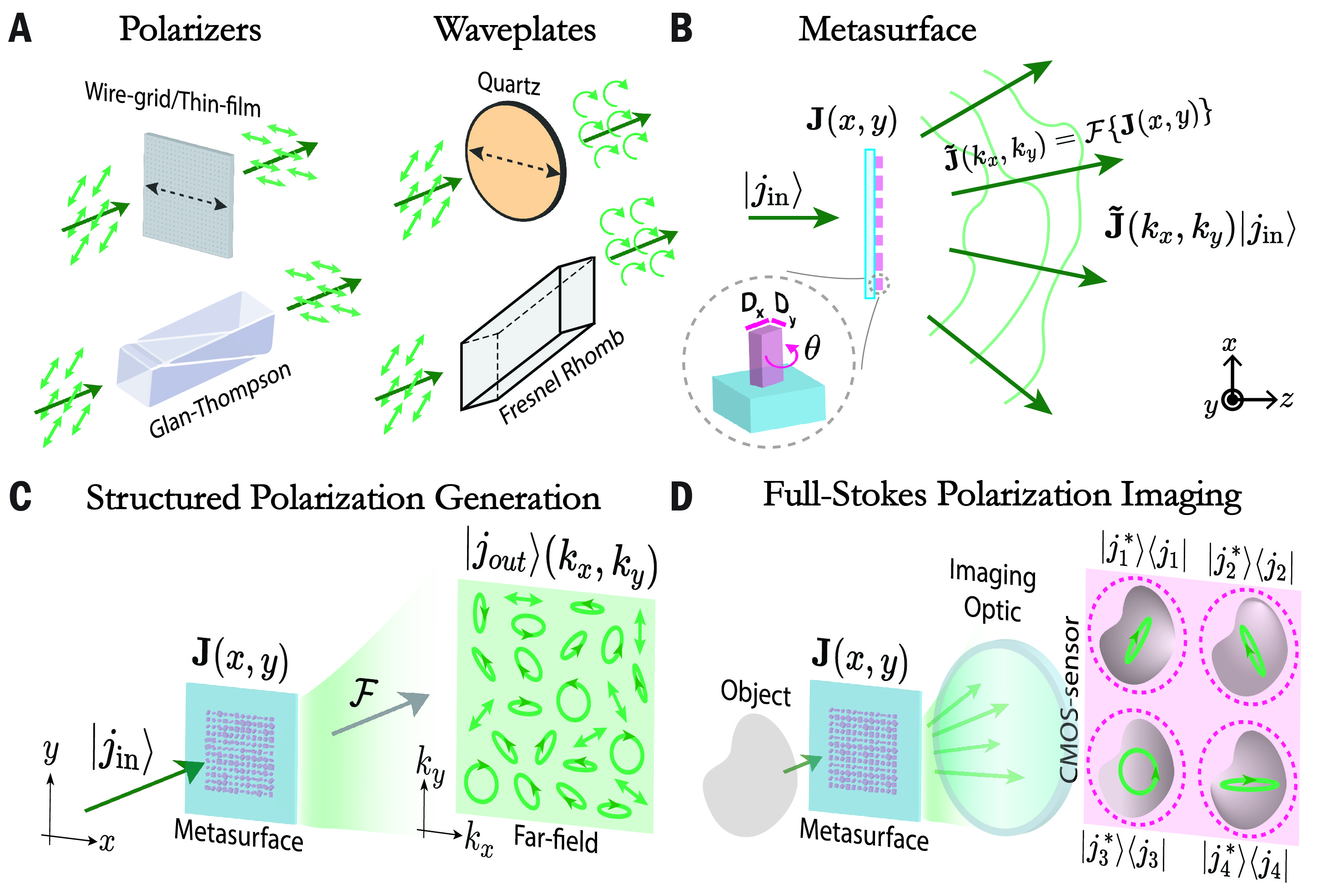}
    \caption{\label{fig:fig1}}
\end{figure*}

Conventional or off-the-shelf polarization optics -- such as wire-grid/thin-film polarizers and quartz crystal waveplates -- transform the incident polarization owing to the bulk polarization response of their constituent elements. These components, while popular, only allow a limited set of polarization transformations, and also provide no spatial control over the output polarization (Fig. \ref{fig:fig1}A). Relatively recent technologies, such as micro-patterning \cite{Andreou2002} and liquid crystals \cite{Andrienko2018}, do provide spatial polarization control, but have their own sets of limitations: micro-patterning is usually limited to linear-only polarization transformations, and has poor spatial resolution \cite{Ratliff2009}; liquid crystals often require multiple passes to address different states of polarization, needing space and introducing complexity \cite{Davis2016}. As a result, MM imaging implementations relying on these components suffer from limitations. For instance, the most popular MM imaging implementation uses the `division-of-time' approach where the sample is sequentially illuminated with and analyzed for different polarization states, in order to compute the entire MM \cite{Berezhnyy2004}. This approach, because of its finite time resolution, is unfeasible for real-time imaging applications. Such an implementation also requires the use of multiple components and moving parts, adding to the bulk, cost and complexity of the system. To mitigate these challenges, in particular the problem of time resolution, researchers have explored partial MM imaging implementations (where only a select few components of the MM are computed) \cite{Gonzalez2020}, and hybrid implementations (consisting a mix of spatial and temporal control) \cite{Angelo2019,Novikova2022} , but a truly single-shot and complete MM imaging implementation has remained elusive. Theoretical and conceptual frameworks of single-shot MM imaging do exist in literature \cite{Kudenov2012,Wang2015,Zaidi2022,Cao2023}, but no convincing implementation has been proposed to date. 

Unlike conventional polarization optics, metasurfaces \cite{Yu2011,Yu2014} -- subwavelength-spaced arrays of phase-shifting nanopillars -- provide complete spatial control over polarization transformations, and outperform technologies such as micro-patterning and liquid crystals, in resolution, form-factor, efficiency and multi-functionality. A metasurface can be mathematically described as a spatially varying Jones matrix of the form 
\begin{equation}
\label{Eq:jones_matrix}
\boldsymbol{J}(x,y) = \boldsymbol{R}(-\theta(x,y)) \begin{bmatrix} e^{i\phi_X(x,y)} & 0 \\ 0 & e^{i\phi_Y(x,y)} \end{bmatrix} \boldsymbol{R}(\theta(x,y))
\end{equation}
where $(x,y)$ are spatial coordinates, and $\boldsymbol{R}(\theta)$ is the $2\times2$ rotation matrix \cite{BalthasarMueller2017}. The three degrees of freedom at each $(x,y)$ -- $\theta(x,y)$, $\phi_X(x,y)$, $\phi_Y(x,y)$ -- are directly related to the shape, including the length ($\text{D}_\text{x}$), width ($\text{D}_\text{y}$),  orientation ($\theta$), and height (kept constant) of the local nanopillar (Fig. \ref{fig:fig1}B) exhibiting strong structural birefringence \cite{Arbabi2015}. As seen in Eq. \ref{Eq:jones_matrix}, a metasurface locally performs unitary transformations only, but through interference, its far-field response can be both unitary (waveplate-like) and/or Hermitian (polarizer-like) \cite{Zaidi2021}. Engineering the far-field -- a region starting hundreds of wavelength away from the aperture and extending all the way to infinity -- to perform user specified polarization transformations is of great utility because of its ease of access, for instance by choosing a suitable operating distance or using a lens. For a metasurface with a spatially varying Jones matrix $\boldsymbol{J}(x,y)$, the polarization response in the far-field (Fig. \ref{fig:fig1}B) can be most succinctly described as $\boldsymbol{\tilde{J}}(k_x,k_y) = \mathcal{F}\{\boldsymbol{J}(x, y)\}$, using matrix Fourier optics \cite{Rubin2019}, where $\mathcal{F}$ is the 2D spatial Fourier transform operator. We can then optimize a metasurface profile $\boldsymbol{J}(x,y)$, using algorithms such as gradient descent or phase retrieval \cite{Fienup1982}, to perform any conceivable polarization transformations $\boldsymbol{\tilde{J}}(k_x,k_y)$ in the far-field. A metasurface can be designed to, as shown in Fig. \ref{fig:fig1}C, generate user-specified spatially varying states of polarization for a given incident polarization \cite{Rubin2018,Arbabi2019,Rubin2021}, or as shown in Fig. \ref{fig:fig1}D, analyze for different polarization states in separate diffraction orders for full-Stokes imaging \cite{Rubin2019,Rubin2022}. Due to their versatility, metasurfaces provide an ideal platform for compact, single-shot and complete MM imaging.

\subsection*{Single-Shot and Complete Mueller Matrix Imaging -- Theory and Design}

\begin{figure*}[t!]
	\centering
    \includegraphics[width=\textwidth]{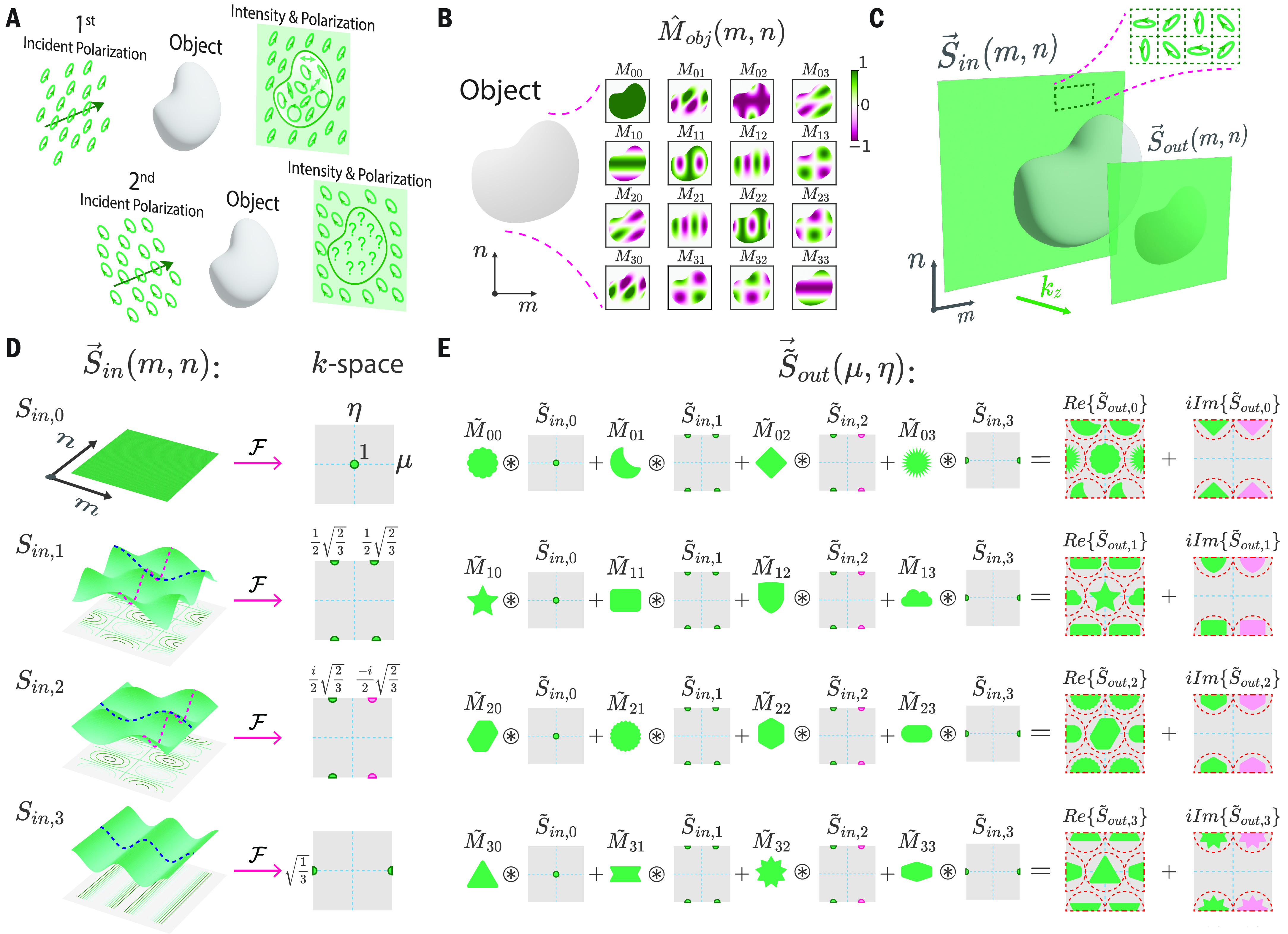}
    \caption{\label{fig:fig2}}
\end{figure*}

Traditionally speaking, a photograph or an image of a non-luminous object describes the spatially varying intensity distribution of transmitted or reflected light under some external illumination. Full-Stokes imaging, of the type shown in Fig. \ref{fig:fig1}D, can describe both the intensity and polarization distribution over the entire field of view -- but the resulting `polarization image' is still incomplete in describing the polarization properties of an object of interest. An object may transmit or reflect a certain distribution of intensity and polarization, given an incident polarization (distribution); however, it might produce a completely different intensity and polarization response under a different set of incident polarizations (Fig. \ref{fig:fig2}A). A complete `polarization image' is thus an image that can describe the response of an object to not only a given incident polarization, but to all possible incident polarizations -- a $4 \times 4$ MM image  (Fig. \ref{fig:fig2}B). An object that may appear invisible or ordinary under standard intensity only imaging, may in fact show rich contrast and distinct features in its MM image. 

For a truly single-shot and complete MM imaging system, the information to compute all the sixteen images that make up a MM image would need to be acquired in a \textit{single measurement} in time. To do this, we consider illuminating the object with a polarization distribution described by the following spatially varying Stokes vector
\begin{equation} \label{Eq:structured_illum}
\vec{S}_{in}(m,n) = 
\begin{pmatrix}
1 \\
\sqrt{2/3}\cos(0.5 m \pi)\cos( n \pi) \\
\sqrt{2/3}\sin(0.5 m \pi)\cos( n \pi) \\
\sqrt{1/3}\cos(m \pi)  
\end{pmatrix}
\end{equation}
where $m$ and $n$ are discrete spatial coordinates. (Because of the use of digital electronics in both computation and measurement, it is best to describe the formulation in terms of discrete spatial coordinates.) For each discrete value of $m$ and $n$, Eq. \ref{Eq:structured_illum} describes a physical state of polarization. In total, the polarization distribution defined by Eq. \ref{Eq:structured_illum}, consists of four unique polarizations (that make up the vertices of a tetrahedron inscribed within the Poincare sphere \cite{supplemental}) arranged in a $2 \times 4$ repeating unit cell (Fig. \ref{fig:fig2}C). Our choice of this particular structured polarization illumination (Eq. \ref{Eq:structured_illum}) is governed by practical and technological considerations \cite{supplemental,Alenin2018}. From Eq. \ref{Eq:structured_illum} we see that the individual Stokes components of the structured polarization illumination, are sinusoids in the spatial domain, and delta functions in the Fourier domain or `$k$-space' (Fig. \ref{fig:fig2}D). Let us now consider an object whose polarization properties are described by its spatially varying Mueller matrix
\begin{equation} \label{Eq:M_obj}
\hat{M}_{obj}(m,n) = 
\begin{pmatrix}
M_{00}(m,n) & M_{01}(m,n) & M_{02}(m,n) & M_{03}(m,n) \\
M_{10}(m,n) & M_{11}(m,n) & M_{12}(m,n) & M_{13}(m,n)  \\
M_{20}(m,n) & M_{21}(m,n) & M_{22}(m,n) & M_{23}(m,n)  \\
M_{30}(m,n) & M_{31}(m,n) & M_{32}(m,n) & M_{33}(m,n)  
\end{pmatrix} 
\end{equation}
When the structured polarization illumination (Eq.\ref{Eq:structured_illum}) interacts with the object (Eq.\ref{Eq:M_obj}), the resulting output Stokes vector can be described as
\begin{equation} \label{Eq:S_out}
\resizebox{0.90\hsize}{!}{$
\vec{S}_{out}(m,n) = 
\begin{pmatrix}
M_{00} + M_{01}\sqrt{2/3}\cos(0.5 m \pi)\cos(n \pi) + M_{02}\sqrt{2/3}\sin(0.5 m \pi)\cos(n \pi) + M_{03}\sqrt{1/3}\cos(m \pi) \\
M_{10} + M_{11}\sqrt{2/3}\cos(0.5 m \pi)\cos(n \pi) + M_{12}\sqrt{2/3}\sin(0.5 m \pi)\cos(n \pi) + M_{13}\sqrt{1/3}\cos(m \pi) \\
M_{20} + M_{21}\sqrt{2/3}\cos(0.5 m \pi)\cos(n \pi) + M_{22}\sqrt{2/3}\sin(0.5 m \pi)\cos(n \pi) + M_{23}\sqrt{1/3}\cos(m \pi) \\
M_{30} + M_{31}\sqrt{2/3}\cos(0.5 m \pi)\cos(n \pi) + M_{32}\sqrt{2/3}\sin(0.5 m \pi)\cos(n \pi) + M_{33}\sqrt{1/3}\cos(m \pi)  
\end{pmatrix}
$}
\end{equation}
 where we omit the $(m,n)$ dependence of the Mueller components for brevity. From Eq. \ref{Eq:S_out}, we can see that the amplitudes of the different spatial \textit{carrier waves} given by the Stokes elements defined in  Eq. \ref{Eq:structured_illum}, are being modulated independently by the different MM components. This \textit{amplitude modulation} can be best understood in $k$-space, by studying the Fourier spectra $\vec{\tilde{S}}_{out}(\mu,\eta)$  -- where $(\mu$,$\eta)$ are the $k$-space coordinates, analogues to spatial coordinates $(m,n)$ -- of the output Stokes vector, defined in terms of convolutions involving the Fourier spectra of the MM components, and the incident Stokes elements. As seen in Fig. \ref{fig:fig2}E, the spectra of every MM component is shifted (by the Fourier shift theorem) to separate locations or `channels' in $k$-space, where these channels are centered on the delta functions given by the spectra of the incident Stokes elements. To avoid any overlap between neighboring spectra of the MM components (aliasing), the separation between two neighboring channels needs to be at least twice the maximum spatial frequency of the spectra of the MM components centered on those channels (Nyquist theorem). If the spectra of each MM component is indeed band-limited and within the Nyquist limit, the image of each MM component can be retrieved without any loss of information. This can be accomplished by the \textit{amplitude demodulation} of the signal(s) in $\vec{S}_{out}$ (Eq. \ref{Eq:S_out}). Thus, if $\vec{S}_{out}$ can be acquired in a single measurement, all the MM components can also be computed in a single measurement -- resulting in a truly single-shot and complete MM imaging system.  

 \subsection*{Experimental Implementation using Metasurfaces}

There are two key modules in our single-shot and complete MM imaging system: structured polarization illumination (using `metasurface 1'), and full-Stokes imaging (using `metasurface 2'). Metasurface 1 generates $\vec{S}_{in}(m,n)$ (Eq. \ref{Eq:structured_illum}), and metasurface 2 analyzes $\vec{S}_{out}(m,n)$ (Eq. \ref{Eq:S_out}) in a single-shot, from which the MM image, $\hat{M}_{obj}(m,n)$ (Eq. \ref{Eq:M_obj}), is computed using the principles defined above. Metasurface 1 and metasurface 2, having separate functions, are computationally optimized using different algorithms \cite{supplemental}; however, they are fabricated using the exact same method \cite{Devlin2016}. The fabricated metasurfaces are composed of TiO$_2$ nanopillars, designed to work with our wavelength of choice (green -- 532 nm) in the visible spectrum. (It should be noted however that the design principles in our work are wavelength agnostic.) Metasurface 1 is composed of an aperiodic arrangement of roughly 4000 nanopillars, with each pillar separated from another by a sub-wavelength distance of 420 nm, making it roughly 1.68 mm in diameter. The structured polarization illumination or `hologram' (with a divergence angle $\theta_{\text{div}}$ of $\pm 40^{\circ}$) diverges quickly, and it is best, in practice, to use a converging lens to access its far-field. Metasurface 2 consists of a periodic arrangement of $12\times12$ arrays of nanopillars (again, separated by 420 nm) and is approximately 3 mm in diameter. Metasurface 2, with a  larger diameter size, allows for more light to pass through the aperture for imaging. 

\begin{figure*}[t!]
	\centering
    \includegraphics[width=\textwidth]{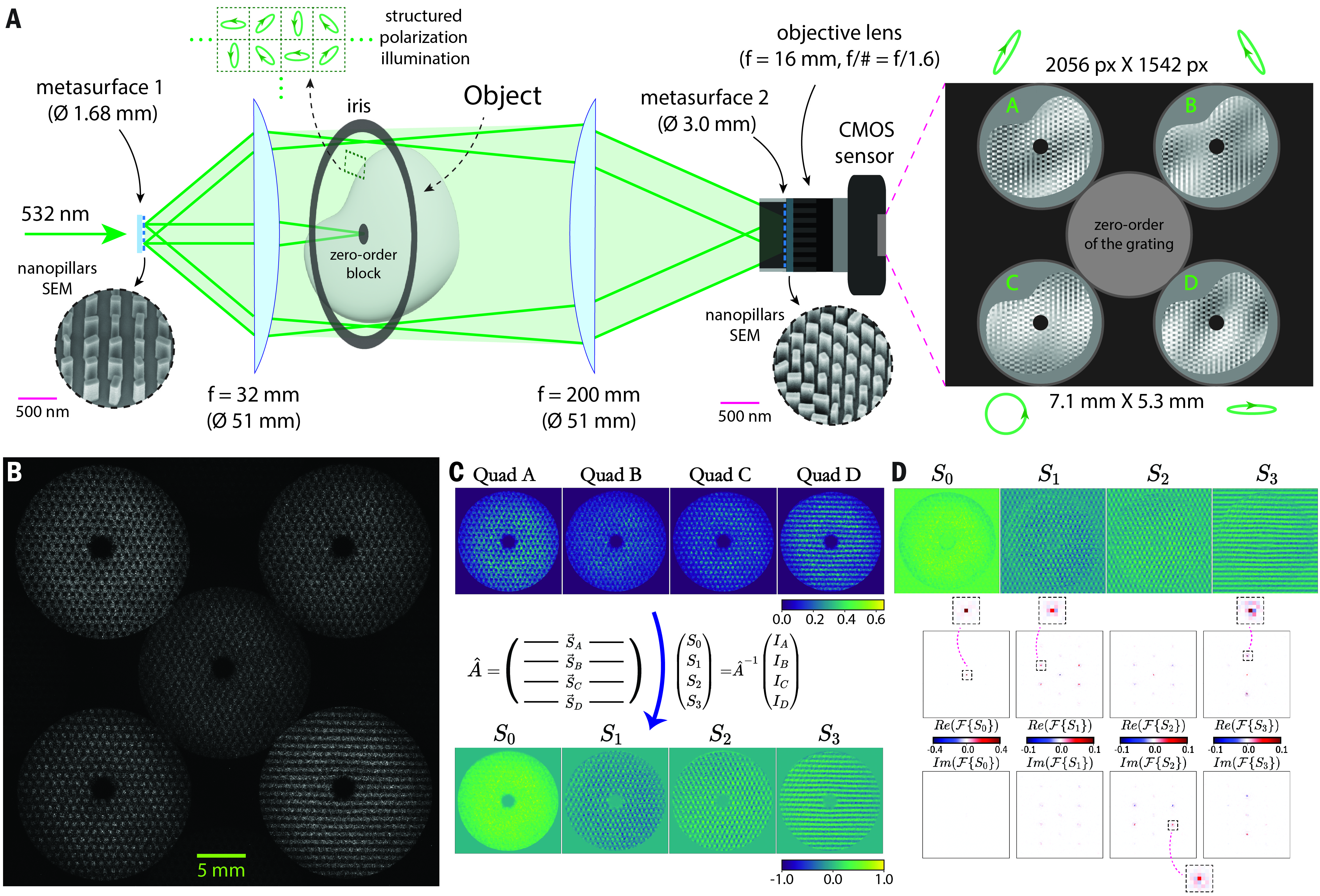}
    \caption{\label{fig:fig3}}
\end{figure*}

In our proof-of-concept implementation (Fig. \ref{fig:fig3}A), we use a $4f$ imaging system to give us control over the spatial and angular (de)magnification of the fields, to completely image the object onto the spatial extent of the CMOS-sensor, but its use is not fundamental to our design. The sizes/dimensions of the metasurfaces, CMOS-sensor, the imaging optic, and the distances between them, can be chosen to eliminate the need of a $4f$ system, depending on the application \cite{supplemental}. The lenses in the $4f$ system can also be replaced by metalenses \cite{Khorasaninejad2016}, for a truly flat optic implementation. The object is placed in the `Fourier-plane', conjugate to both metasurface 1 and metasurface 2. An iris and a zero-order beam block (circular black Acktar tape on a glass substrate) are also placed in the Fourier plane to limit the field-of-view (FOV) and block the background zeroth order light, respectively. The object, now in the far-field of metasurface 1, interacts with the structured polarization illumination; the resulting fields, in transmission and/or in reflection, are then imaged by the full-Stokes camera to retrieve the complete MM image.

The reconstruction accuracy of the MM image is directly related to the accuracy of the full-Stokes camera. It is important, thus, to independently design and calibrate a robust metasurface full-Stokes camera, as we have previously demonstrated \cite{Rubin2022,Li2023}, before introducing it into our MM imaging setup. Our metasurface full-Stokes camera consists of metasurface 2, an imaging optic, and a CMOS-sensor. Metasurface 2 functions as both a diffraction grating (to `split amplitudes'), and an analyzer to simultaneously analyze for polarization states in different diffraction orders. In particular, metasurface 2 analyzes for four different polarization states (that -- again -- make up the vertices of a tetrahedron inscribed within the Poincare sphere \cite{supplemental}), in the first four off-axis diffraction orders. When a scene, extended over some FOV, is incident on metasurface 2, the separately analyzed copies of the scene in the four diffraction orders can be imaged onto a CMOS-sensor using an imaging optic. The four-element Stokes vector necessitates at least four measurements to be fully determined (at each pixel), hence the use of four diffraction orders (and not any less). If the metasurface grating is designed to analyze for the four polarization states $\vec{S_A},\vec{S_B},\vec{S_C},\vec{S_D}$, then given a incident spatially varying Stokes vector $\vec{S}_{out}$, the 4-element `intensity vector' (a vector of images) $\vec{I}_{out}(m,n)$ can be written as 
\begin{equation} \label{Eq:int_vec}
\overbrace{\begin{pmatrix}
I_{out,A}(m,n) \\
I_{out,B}(m,n)\\
I_{out,C}(m,n)\\
I_{out,D}(m,n)   
\end{pmatrix}}^{\vec{I}_{out}(m,n)}
=
\overbrace{\begin{pmatrix}
\text{------} \vec{S}_A \text{------}\\
\text{------} \vec{S}_B \text{------}\\
\text{------} \vec{S}_C \text{------}\\
\text{------} \vec{S}_D \text{------}   
\end{pmatrix}}^{\hat{A}}
\overbrace{\begin{pmatrix}
S_{out,0}(m,n)\\
S_{out,1}(m,n)\\
S_{out,2}(m,n)\\
S_{out,3}(m,n)
\end{pmatrix}}^{\vec{S}_{out}(m,n)}
\end{equation}
where $\hat{A}$ is referred to as the instrument matrix. Using Eq. \ref{Eq:int_vec}, $\vec{S}_{out}$ can be determined -- pixel by pixel at each $(m,n)$ over the entire FOV -- by using the inverse of the instrument matrix, $\hat{A}^{-1}$. The instrument matrix $\hat{A}$ is experimentally determined through a thorough calibration \cite{Rubin2022,Li2023}, to calibrate out discrepancies between the ideal design, and its practical implementation by metasurface 2.

Fig. \ref{fig:fig3}B shows the raw grayscale image output by the system, in the case of no object, or simply `air'. We can see the circular edges of the iris and the circular beam block. The iris, functioning as a field stop, limits the FOV and prevents the scenes centered on different diffraction orders from overlapping with each other. The zeroth order beam block prevents the strong background laser light from propagating through the system, and saturating the sensor. The raw image can be processed into a full-Stokes image using the instrument matrix $\hat{A}$ (Fig. \ref{fig:fig3}C). We can then compute the complete MM image through amplitude demodulation of the full-Stokes image. There are different methods/algorithms to demodulate a signal from a carrier-wave \cite{Oppenheim1996-sa}; our method of choice is the `product detector' method. In this method, the amplitude modulated signal is multiplied by the carrier wave, followed by a low pass filter \cite{supplemental}. This method requires the knowledge of the carrier wave, which necessitates a reference measurement. In our case, the `air' image serves as the reference measurement, since air has a spatially uniform and identity MM. The reference measurement also helps us calibrate out any spatial variations in the overall intensity of our structured polarization illumination. To prepare the air Stokes image as a reference measurement to be used in demodulation, we complete parts of the image with no signal (corresponding to the regions where light is blocked from the iris and the beam block), through extrapolation (Fig. \ref{fig:fig3}D). This ensures that the reference Stokes components are close to the desired sinusoid signals. This can be seen in the Fourier spectra of the reference Stokes components -- we see (approximate) delta functions in the expected channel locations (Fig. \ref{fig:fig3}D). 

\subsection*{Measurement Results -- In Transmission and Reflection}

\begin{figure*}[t!]
	\centering
    \includegraphics[width=\textwidth]{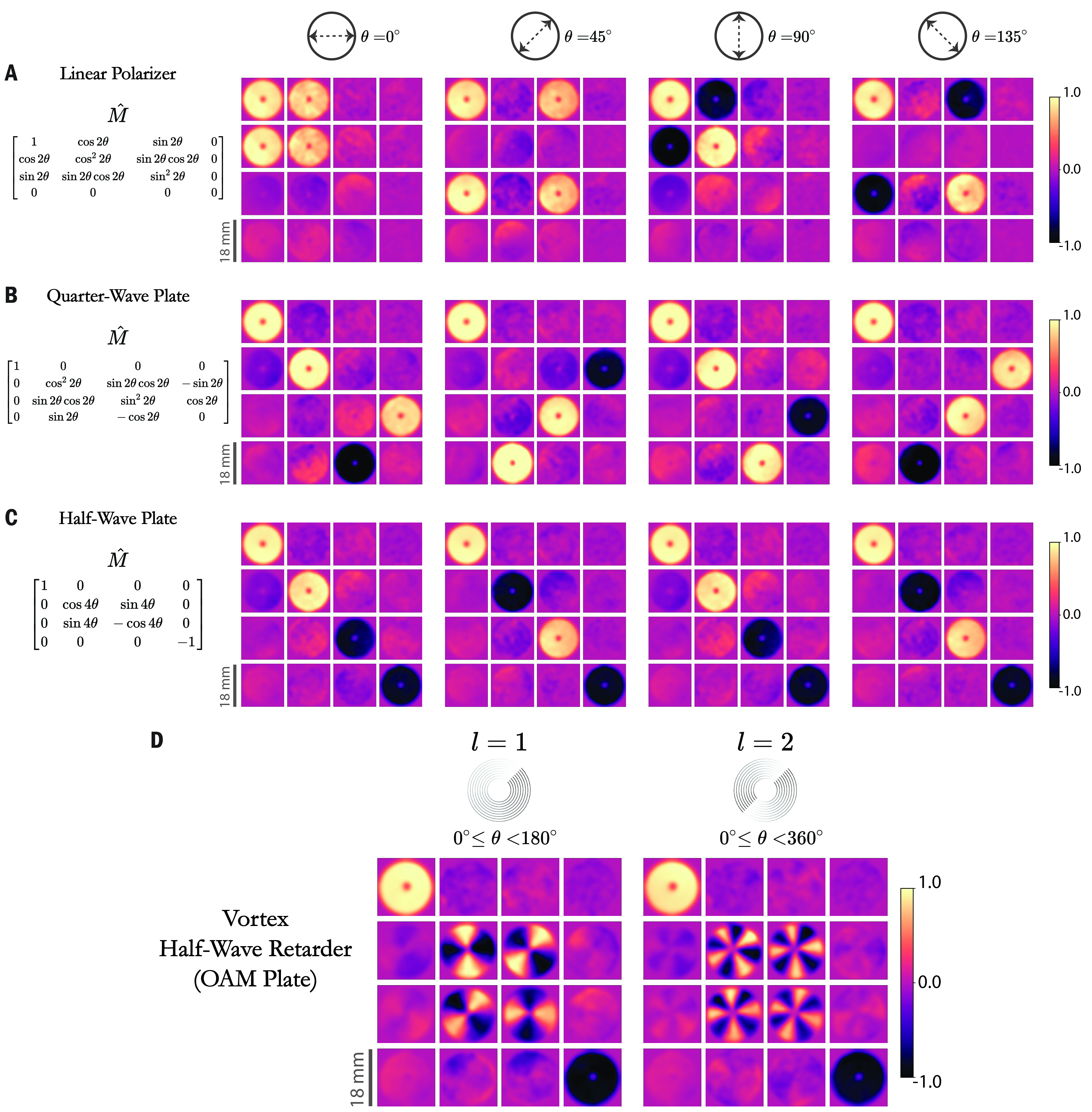}
    \caption{\label{fig:fig4}}
\end{figure*}

We use our system to image a variety of commercially available polarization optics, such as polarizers, waveplates and orbital-angular-momentum (OAM) plates. The results are summarized in Fig. \ref{fig:fig4}. We image a linear film polarizer (Fig. \ref{fig:fig4}A), a zero-order quarter-waveplate (QWP) (Fig. \ref{fig:fig4}B), and a zero-order half-waveplate (HWP) (Fig. \ref{fig:fig4}C), each in four different orientations, resulting in twelve different MM images. The expected (theoretical) MM responses of these optics as a function of different orientations ($\theta$), are also listed in Fig. \ref{fig:fig4}A-C. We see good correspondence between the measured MM images and the expected responses. Furthermore, we can easily identify the regions where the light is blocked by the iris and the beam block, having MM values of zero. The edges of the iris and the circular beam block, in a way, provide us with additional spatial information to image and test our system beyond just the `polarization active' region of the optics. A MM is often normalized with respect to its $M_{00}$ component. For a MM image, normalizing each pixel with respect to its $M_{00}$ component will result in the loss of the overall (polarization-insensitive) intensity variation across the image. Therefore, we normalize each pixel of all the components of the MM image, by the maximum pixel value of the $M_{00}$ image. (Following normalization, saturated pixels, with nonphysical values beyond $\pm 1$, are clipped at $\pm 1$. Saturated pixels, still, have values very close to $\pm 1$, as seen in our analysis \cite{supplemental}.) We also image two OAM plates, results of which are shown in Fig. \ref{fig:fig4}D. In particular, these are vortex half-wave retarders\footnote{\url{https://www.thorlabs.com/newgrouppage9.cfm?objectgroup_id=9098}} made of liquid crystal polymers, with a fixed retardance of $\pi$, but a fast-axis that rotates azimuthally over the area of the optic. This spatial variation is unambiguously captured in the $M_{11},M_{12},M_{21}$ and $M_{22}$ components in the two Mueller images shown in Fig. \ref{fig:fig4}D.

As explained above, we calibrate the full-Stokes camera and perform reference measurements for the structured polarization illumination -- still, inevitably, there remain sources of error that can degrade the quality of the final image. Contributions from high spatial frequency components from, for instance, the sharp edges of the iris and the beam block, can result in aliasing, though empirically, we find these contributions to be negligible \cite{supplemental}. A main source of error, it would seem, is the channel `cross-talk'. The Stokes components defined in Eq. \ref{Eq:structured_illum} are ideally all orthogonal to each other, but in practice, when implemented by the metasurface, the resulting reference Stokes components (Fig. \ref{fig:fig3}D) are not completely orthogonal resulting in (small) mixing of the spectra of different Mueller components, or cross-talk.

\begin{figure*}[t!]
	\centering
    \includegraphics[width=\textwidth]{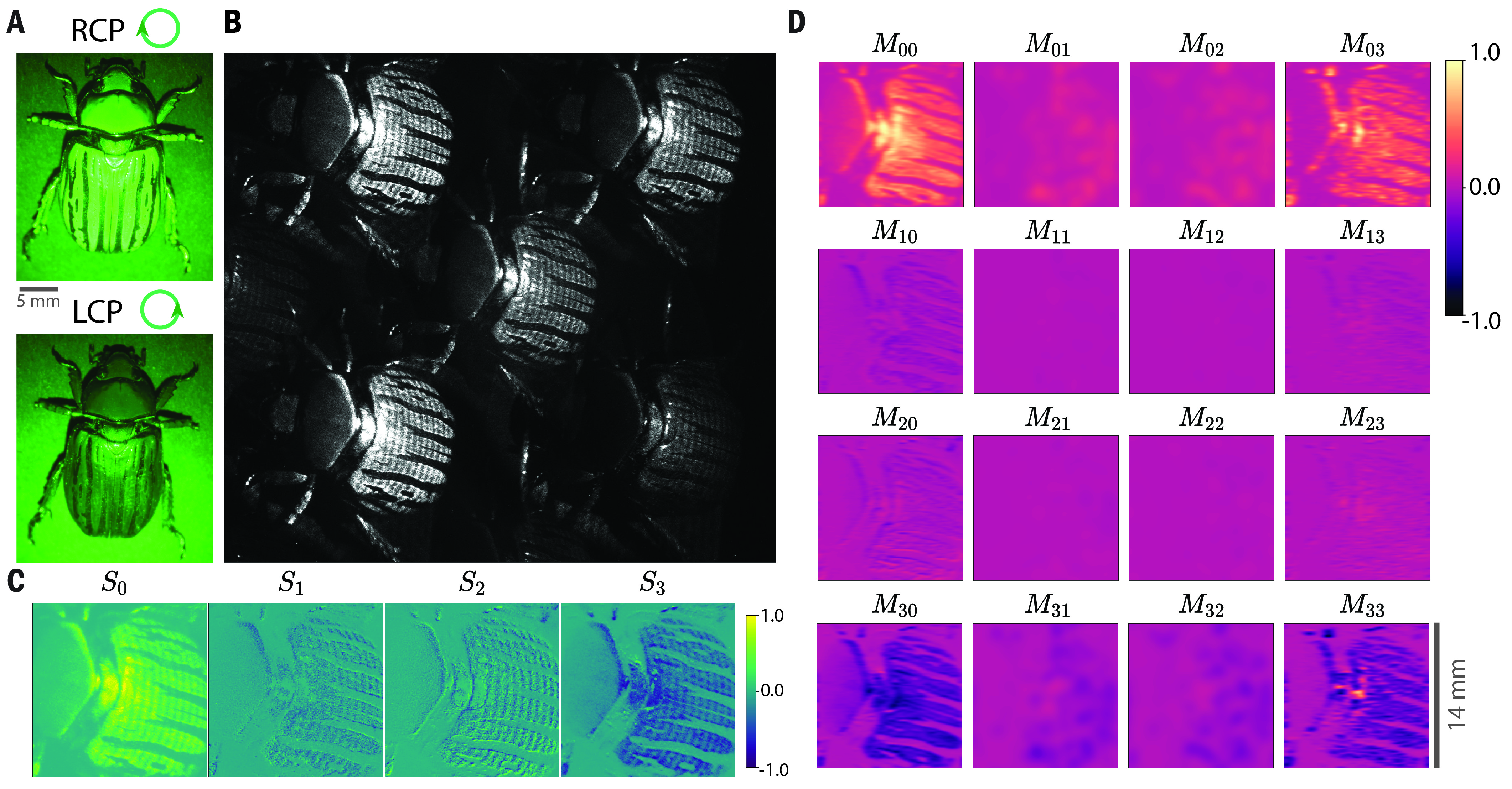}
    \caption{\label{fig:fig5}}
\end{figure*}

We also test our system for imaging in reflection. We decided to image \textit{Chrysina gloriosa}, or the `chiral beetle', which is known to have different optical responses for the two circular polarizations. As shown in Fig. \ref{fig:fig5}A, parts of the exoskeleton or shell of the chiral beetle appear bright under right-circular polarization (RCP) illumination and dark under left-circular polarization (LCP) illumination (for the green wavelength). Thus, we can ascertain that the shell of the chiral beetle is analyzing for circular polarization; still, we cannot conclude the state-of-polarization reflecting off of the shell from these two intensity measurements. Our MM imaging system, set up in reflection, allows us to determine the exact polarization properties of the shell of the chiral beetle (Fig. \ref{fig:fig5}B-D). We could not determine a suitable reflective object for reference Stokes measurements in reflection, due to the curvature of the beetle's shell as evident in the raw image (Fig. \ref{fig:fig5}B). Therefore, to compute the MM image (Fig. \ref{fig:fig5}D), from the Stokes image (Fig. \ref{fig:fig5}C), we use a reference free method, by using contextual knowledge and some simplifying assumptions \cite{supplemental}. From the resulting MM image of the chiral beetle (Fig. \ref{fig:fig5}D), we can determine that its shell behaves as a \textit{homogenous circular polarizer}  -- it not only analyzes for circular polarization, but also generates circular polarization in reflection. Interesting spatial features such as the \textit{striae} (characteristic dark lines on the shell of the chiral beetle) are also resolved in the final MM image (Fig. \ref{fig:fig5}D), by using a rectangular low-pass filter (assigning more bandwidth along one axis in comparison to the other) \cite{supplemental}. 

\subsection*{Discussion and Conclusion}

In our work, we have described the principles behind designing and building a compact, single-shot and complete Mueller matrix imaging system, and using those principles, implemented the described system to successfully image objects in both transmission and reflection. In designing our system, we settled on a resolution that was sufficient for our proof-of-concept measurements. By designing larger metasurfaces, using higher NA lenses and higher resolution sensors, the resolution of the MM imaging system can be pushed to its theoretical limit \cite{supplemental}. More sophisticated calibration methods and reconstruction algorithms, for instance by custom designing the Fourier filters \cite{li2021spectral}, can be used to further reduce errors from aliasing and cross-talk. Existing super-resolution techniques, and machine learning assisted reconstruction can also be used to image beyond the band-limits of the filters \cite{Yang2019}. Furthermore, the structured polarization illumination can be used to probe the depth profile of the object in addition to its polarization properties, for \textit{polarization-resolved depth sensing}. (For example, the curvature of the lines reflecting off of the shell of the chiral beetle in Fig. \ref{fig:fig5}B, could be potentially used to estimate its curvature.) These directions will be the subject of future work.

Our work illustrates the ability of metasurfaces to greatly simplify the design and architecture of devices and systems based on polarization optics. The MM imaging system we present and the methods underlying its design should enable the adoption of MM imaging for a myriad of applications, including those that require both compact and real-time polarization imaging. Our work has the potential to not only empower existing MM imaging applications in important fields in biomedicine, most noticeably in cancer detection, but also inspire research in newer directions such as depth-resolved MM confocal microscopy \cite{Lara2006} and polarimetric endscopic imaging \cite{Pahlevaninezhad2018, Qi2023}. The potential of MM imaging in saccharimetry (the process of measuring the amount of sugar in a sample) would be useful to food, pharmaceutical and biomedical industries. MM imaging, fittingly, could also be of great use in the characterization of nanostructures, metasurfaces and metamaterials \cite{Oates2014, Schmidt2014}. Consumer electronics applications, such as facial recognition in smartphones, and eye-tracking in augmented and virtual reality headsets, should also greatly benefit from the small form factor of our system for complete and single-shot MM imaging. Furthermore, this system, given its superior time resolution and flexibility, could be useful in generating large MM datasets to train neural networks for many machine learning classification applications. Beyond technological applications, our work could be of consequence in fundamental science, such as in the detection of the time-varying birefringence of vacuum in the presence of intense electric and magnetic fields (as theorized by quantum electrodynamics) \cite{Heinzl2006}, in the study of 3D polarization states of light \cite{Azzam2011}, and in the research of both short wavelength (X-ray) \cite{Marx2013} and long wavelength (terahertz) polarimetry \cite{Wiesauer2013}. This should inspire research in exciting new directions.

% Your references go at thea particular  end of the main text, and before the
% figures.  For this document we've used BibTeX, the .bib file
% scibib.bib, and the .bst file Science.bst.  The package scicite.sty
% was included to format the reference numbers according to *Science*
% style.

%BibTeX users: After compilation, comment out the following two lines and paste in
% the generated .bbl file. 

\bibliography{scibib}

\bibliographystyle{Science}

\section*{Acknowledgments}
This material is based upon work supported by the Air Force Office of Scientific Research under award
Number FA9550-21-1-0312.

%Here you should list the contents of your Supplementary Materials -- below is an example. 
%You should include a list of Supplementary figures, Tables, and any references that appear only in the SM. 
%Note that the reference numbering continues from the main text to the SM.
% In the example below, Refs. 4-10 were cited only in the SM.     
\section*{Supplementary materials}
Supplementary document is available upon request.

% For your review copy (i.e., the file you initially send in for
% evaluation), you can use the {figure} environment and the
% \includegraphics command to stream your figures into the text, placing
% all figures at the end.  For the final, revised manuscript for
% acceptance and production, however, PostScript or other graphics
% should not be streamed into your compliled file.  Instead, set
% captions as simple paragraphs (with a \noindent tag), setting them
% off from the rest of the text with a \clearpage as shown  below, and
% submit figures as separate files according to the Art Department's
% instructions.

\clearpage

\noindent {\bf Fig. 1. Polarization Optics -- Conventional vs. Metasurfaces.} (\textbf{A}) Examples of popular conventional polarization optics, including polarizers and waveplates. Conventional polarization optics, owing to its bulk polarization response, provides no spatial control over the resulting polarization. (\textbf{B}) A metasurface -- which can be described by a spatially varying Jones matrix distribution $\mathbf{J}(x,y)$, and a far-field polarization response $\mathbf{\tilde{J}}(k_x,k_y) = \mathcal{F}\{\mathbf{J}(x, y)\}$, where $\mathcal{F}$ is the 2D spatial Fourier transform, provides full spatial control over the resulting polarization. Each nanopillar within the metasurface provides three degrees of freedom -- pillar length $D_x$, pillar width $D_y$ and pillar orientation $\theta$; the metasurface, thus, made up of thousands of such nanopillars in either a periodic or an aperiodic arrangement, has many degrees of freedom which can be optimized for a user-specified polarization response. (\textbf{C}) A metasurface can be used for `structured polarization' generation, where, for an incident polarization state $\ket{j_{\text{in}}}$, the far-field output $\ket{j_{\text{out}}}(k_x,k_y)$, given by $\mathbf{\tilde{J}}(k_x,k_y)\ket{j_{\text{in}}}$, follows a user-specified polarization distribution. (\textbf{D}) A metasurface can also be designed to function as a polarization analyzer with a Jones-matrix response $\mathbf{\tilde{J}} = \ket{j^*(k_x,k_y)}\bra{j(k_x,k_y)}$, in the far-field. A (periodic) metasurface -- that splits the incoming light into its first four diffraction orders, and simultaneously analyzes for four different polarization states -- alongside an imaging lens and a sensor, can perform full-Stokes imaging.

\clearpage

\noindent {\bf Fig. 2. Single-shot and Complete Mueller Matrix Imaging -- Why and How?} (\textbf{A}) For a given incident polarization, an object, in general, produces a spatially varying intensity and polarization response in transmission or reflection, which can be imaged by, for instance, a full-Stokes camera. For a different incident polarization, the object may produce a completely different intensity and polarization response. (\textbf{B}) To fully quantify the spatially varying polarization properties of an object, we have to calculate a set of 16 images that correspond to its $4 \times 4$ MM image. An object can reveal rich contrast in different components of its MM image, making it much easier to classify. (\textbf{C}) We can spatially sample the object with a minimum of four different incident polarization states $\vec{S}_{in}(m,n)$, where the resulting polarization distribution $\vec{S}_{out}(m,n) = \hat{M}_{obj}(m,n)\vec{S}_{in}(m,n)$ can be imaged by a full-Stokes camera. (\textbf{D}) The different Stokes components that up make the incident polarization states, are spatially varying and orthogonal sinusoids with different spatial frequencies and thus non-overlapping delta functions in the Fourier-domain. (\textbf{E}) $\vec{S}_{out}(m,n) = \hat{M}_{obj}(m,n)\vec{S}_{in}(m,n)$, instead of (matrix) multiplication, can be described in terms of convolutions ($\circledast$) by describing the computation in the Fourier domain. We see that the Fourier spectra of the different components of the object's MM, $\hat{M}_{obj}(m,n)$, `shift' to the location of the different delta functions corresponding to the spectra of the different incidents Stokes components -- due to the convolution operation. If the spectra of the Mueller components are sufficiently band-limited, they can be `filtered' from the resulting Stokes image, allowing the reconstruction of $\hat{M}_{obj}(m,n)$ without any loss of information.

\clearpage

\noindent {\bf Fig. 3. Experimental Implementation and Calibration.} (\textbf{A}) A $4f$ imaging system is used to image the MM of the object, which is placed in the `Fourier plane', conjugate to both metasurface 1 and metasurface 2. Metasurface 1 produces structured polarized light which illuminates the object, and metasurface 2 diffracts and simultaneously analyzes the resulting fields which are then imaged onto the CMOS-sensor. The iris in the Fourier-domain is placed to limit the FOV, and the zero-order block is placed to prevent the strong background laser light from saturating the sensor. (\textbf{B}) Raw measured image produced by the system in the case of no object. The periodic checkerboard and grid-like patterns correspond to the spatially varying incident polarization. This image of `air', which has an identity Mueller matrix, can be used for calibration. (\textbf{C}) The raw images in the four quadrants of the CMOS-sensor can be converted to the corresponding full-Stokes image by using the instrument matrix, which is predetermined in a separate, full-Stokes calibration. (\textbf{D}) The region with no signal (corresponding to the iris and zeroth order block) can be `filled' via extrapolation -- resulting in images or signals which are close to 2D sinusoids. These images can then be used as reference images for amplitude demodulation. We also plot the Fourier spectra of these reference images which are close to the expected delta-like functions.  

\clearpage

\noindent {\bf Fig. 4. Mueller Matrix Imaging Results - In Transmission.} (\textbf{A}) Measured MM images of a thin-film linear polarizer at four different transmission-axis orientations ($\theta$). Also listed is the MM ($\hat{M}$) of the linear polarizer as a function of $\theta$. (\textbf{B}) Measured MM images of a zero-order quartz quarter-wave plate at four different fast-axis orientations ($\theta$). Also listed is the MM ($\hat{M}$) of the quarter-wave plate as a function of $\theta$. (\textbf{C}) Measured MM images of a zero-order quartz half-wave plate at four different fast-axis orientations ($\theta$). Also listed is the MM ($\hat{M}$) of the half-wave plate as a function of $\theta$. (\textbf{D}) Measured MM images of two, $l=1$ and $l=2$, vortex half-wave retarders (commonly known as OAM plates).

\clearpage

\noindent {\bf Fig. 5. Mueller Matrix Imaging Results - In Reflection.} (\textbf{A}) \textit{Chrysina gloriosa}, more commonly referred to as the `chiral beetle', is illuminated by right circularly polarized (RCP) light and left circularly polarized (LCP) light, and imaged by a standard digital camera. The intensity images, juxtaposed for comparison, show that the beetle exhibits a different optical response for the two circular polarizations. (\textbf{B}) Raw image of the chiral beetle captured using our MM imaging system. (\textbf{C}) The resulting full-Stokes image computed from the raw image, using the instrument matrix $\hat{A}$ (Eq. \ref{Eq:int_vec}). (\textbf{D}) The resulting MM image from the full-Stokes image, is computed (demodulated and normalized) using a reference free method \cite{supplemental}. The MM image of the chiral beetle, has spatially resolved features such as the size/shape of the shell, the characteristic \textit{striae} (or lines) on the shell; it also confirms that the shell of the chiral beetle functions as a homogeneous circular polarizer.

\end{document}